**Carbon Layer Orientation and Closed-Pore Construction Achieving Ultra-Low Specific Surface Area Hard Carbon for High-Performance Na-ion Storage**

*Bowen Wang[a], Zihan Yang[b], Minghui Zhao[a], Wenjie Mai[b], Qing Xu[a], Huan Li[a]*, Liang Zhang[d], Chul Gyu Jhun[d], Le Chen[c]*, Wentao Zhang[c], Jingtai Zhao[a], Jinliang Li[b]**

[a]*School of Materials Science and Engineering, Guilin University of Electronic Technology, Guilin 541004, China*

[b]*Siyuan Laboratory, Guangdong Provincial Engineering Technology Research Center of Vacuum Coating Technologies and New Materials, Guangdong Provincial Key Laboratory of Nanophotonic Manipulation, Department of Physics, College of Physics & Optoelectronic Engineering, Jinan University, Guangzhou 510632, China*

[c]*Guangxi Key Laboratory of Optoelectronic Information Processing, School of Optoelectronic Engineering, Guilin University of Electronic Technology, Guilin 541004, China*

[d]*Department of Semiconductor Engineering, Hoseo University, Asan 31499, Korea*

**Corresponding authors: lijinliang@email.jnu.edu.cn (J. Li); chenle11@126.com (L. Chen); lihuan@guet.edu.cn (H. Li)*

**Abstract**

Addressing the critical trade-off between initial Coulombic efficiency (ICE) and reversible capacity in hard carbon anodes for Na-ion batteries (NIBs), we introduce a novel coupling strategy that combines carbon layer orientation reconstruction with closed-pore construction to produce hard carbon with an ultra-low specific surface area. We demonstrate that the nanographite domains within the hard carbon precursor undergo entropy-driven orientation reconstruction through the synergistic regulation of heteroatom doping and medium-temperature carbonization. This process not only increases interlayer spacing and promotes structural disorder but also enables the formation of dense, closed pores and ultramicropores at domain boundaries via confined atomic migration, while simultaneously encapsulating surface open pores within internal closed ones. Due to this unique pore architecture, our hard carbon exhibits an ultra-low specific surface area of 1.89 $m^2$ $g^{-1}$ with a markedly higher proportion of closed pores. As a result, our hard carbon achieves a remarkable reversible capacity of 342.3 mAh $g^{-1}$ at 20 mA $g^{-1}$, with an exceptional ICE of 90.4% and a dominant plateau capacity of 262.3 mAh $g^{-1}$ (76.6%) for NIBs. We believe this coupling strategy provides a new paradigm for the structural engineering of high-ICE anode materials in advanced NIBs.

## 1. Introduction

Na-ion batteries (NIBs) are regarded as a pivotal supplement to lithium-ion batteries, thanks to the plentiful reserves and low cost of sodium.[1] In NIBs, the choice of anode material critically impacts electrochemical performance. Hard carbon has attracted considerable attention due to its low operating potential, adjustable interlayer spacing, and strong chemical stability.[2-4] However, commercial use of hard carbon anodes faces two main challenges: low initial Coulombic efficiency (ICE), which reduces energy density, and limited plateau capacity, which impacts stability.[5] These issues mainly stem from difficulty in controlling the microscopic pore structure; a high specific surface area increases parasitic side reactions, while a lack of closed pores limits plateau capacity. These factors hinder the widespread adoption of HC-based NIBs.

For hard carbon, a low ICE indicates that a substantial amount of active Na from the cathode is irreversibly consumed on the anode surface and in the formation of the solid electrolyte interphase (SEI) layer during the initial sodiation.[6] This portion of Na cannot contribute to capacity in subsequent cycles, thereby directly sacrificing the energy density and economic viability of the full battery. In addition, the sodiation-desodiation process of hard carbon consists of a slope region above 0.1 V and a plateau region below this threshold, with the latter contributing significantly to the total energy density. Tracing back to the origin, both the interfacial side reactions involved in irreversible Na loss and the pore-filling behavior relied upon for plateau storage are intimately correlated with the internal microscopic pore structure of the carbon material.[7] An excessively high specific surface area provides abundant reaction sites for electrolyte decomposition, leading to excessive SEI layer growth and a consequent decline in ICE. The magnitude of the plateau capacity depends on the formation of a sufficient number of closed pores or ultramicropores to accommodate Na cluster deposition.[8] Therefore, constructing a dense network of closed pores/ultramicropores while suppressing the specific surface area has become the crux of pore-structure engineering in hard carbon.

Regarding closed-pore construction and carbon layer structure regulation, existing strategies have made progress, but still suffer from inherent limitations. The first approach is the template method, which employs templates such as MgO, $SiO_2$, or salts to introduce abundant pore structures.[9] Although template-based techniques can enhance the plateau capacity of hard carbon, they typically increase production costs and involve acid and base treatments that raise environmental concerns.[10-11] Furthermore, the resulting hard carbon often contains excessive defects and uncontrollable closed pores, leading to a low ICE. The approach involves modifying the chemical vapor deposition (CVD) coating. While CVD can effectively seal open pores with a carbon layer, traditional processes are limited by slow methane decomposition kinetics and lengthy deposition cycles, and overly ordered carbon layers can hinder $Na^+$ transport and storage.[12-13] Additionally, heteroatom doping, such as nitrogen and sulfur, in hard carbon can create defect sites within the carbon framework, thereby boosting slope capacity. However, heteroatom doping is often accompanied by an increased specific surface area and intensified side reactions, resulting in a decreased ICE. More critically, relying solely on heteroatoms makes it difficult to achieve directional construction of a closed-pore network while suppressing the specific surface area, thereby leaving the pore-structure characteristics inadequately regulated.[14-15]

Herein, we propose a coupled regulation strategy integrating carbon layer orientation reconstruction with closed-pore construction. This strategy leverages the local structural distortion induced by heteroatom introduction to provide a thermodynamic driving force for the limited ordered rearrangement of carbon layers during medium-temperature carbonization. We find that the carbon layers undergo moderate reconstruction by precisely regulating the heat-treatment window, thereby converting originally open surface pores into closed pores. Experimental results demonstrate that the obtained hard carbon sample (PHC-800) possesses an ultra-low specific surface area of 1.89 $m^2$ $g^{-1}$, achieving a high ICE of 90.4% and a high reversible capacity of 342.3 mAh $g^{-1}$. The optimized material significantly enhances the ICE while retaining a high reversible capacity, validating the feasibility of this coupling mechanism in overcoming the core bottleneck of reconciling high capacity and high ICE in hard carbon anodes.

**2. Results and Discussion**

We selected coconut shell as the carbon precursor and introduced phosphoric acid to tailor the carbon structure. The larger atomic radius of P in phosphate anions promotes defect formation, such as five- and seven-membered rings, when incorporated, aiding the development of thin pseudo-graphitic layers and improving the electronic conductivity and reaction kinetics of the resulting hard carbon (Fig. 1a).[16-17] Recognizing the importance of porous architecture in Na-ion storage, an intermediate-temperature (MT) heat treatment was employed to enhance P doping. The structural distortions caused by P facilitate partial reconstruction of the carbon framework and atomic migration within the MT regime, which together induce reorientation of the carbon layers. This process gradually transforms some initially open pores into closed pores, forming a dense network of ultramicropores at domain boundaries. For clarity, we define the directly annealed hard carbon, the hard carbon with an MT zone treatment, the hard carbon with added P, and the hard carbon modified with both P and the MT zone treatment as HC, HC-800, PHC, and PHC-800, respectively. Fig. 1b-e presents the TEM images of HC, HC-800, PHC, and PHC-800. It is found that the HC features a relatively dense surface with an interlayer spacing of 0.361 nm and primarily long-range graphitic order (Fig. 1b). After MT treatment, the HC-800 (Fig. 1c) undergoes moderate framework reconstruction, forming short-range ordered pseudo-graphitic microcrystals with an expanded interlayer spacing of 0.372 nm and increased structural disorder. The PHC (Fig. 1d) exhibits a further expanded interlayer spacing of 0.374 nm, although in the absence of MT treatment, the overall ordering remains relatively high. In contrast, PHC-800 (Fig. 1e) displays highly disordered short-range graphitic microcrystals and large closed pores, with an interlayer spacing of 0.377 nm. These observations indicate that the combination of P doping and MT treatment effectively transforms hard carbon from long-range-ordered to short-range-disordered structures. P incorporation increases interlayer spacing, thereby activating the interlayer $Na^+$ insertion mechanism and expanding ionic transport channels. The synergistic effect of MT-induced reconstruction optimizes the carbon layer arrangement, enabling the formation of well-developed closed pores that enhance the filling capacity of Na clusters.[18] SEM imaging further illustrates the morphological evolution of hard carbon under different treatment conditions (Fig. S1). To assess the distribution and incorporation of P in PHC-800, FIB-SEM was used to section the sample and perform elemental mapping. As shown in Fig. 1f, PHC-800 exhibits loosely packed particles with well-developed internal porosity, promoting electrolyte infiltration and ion transport. Elemental mapping of the sectioned regions (Fig. 1g-i and Table S1) confirms uniform P distribution, indicating effective, homogeneous incorporation into the carbon layers.

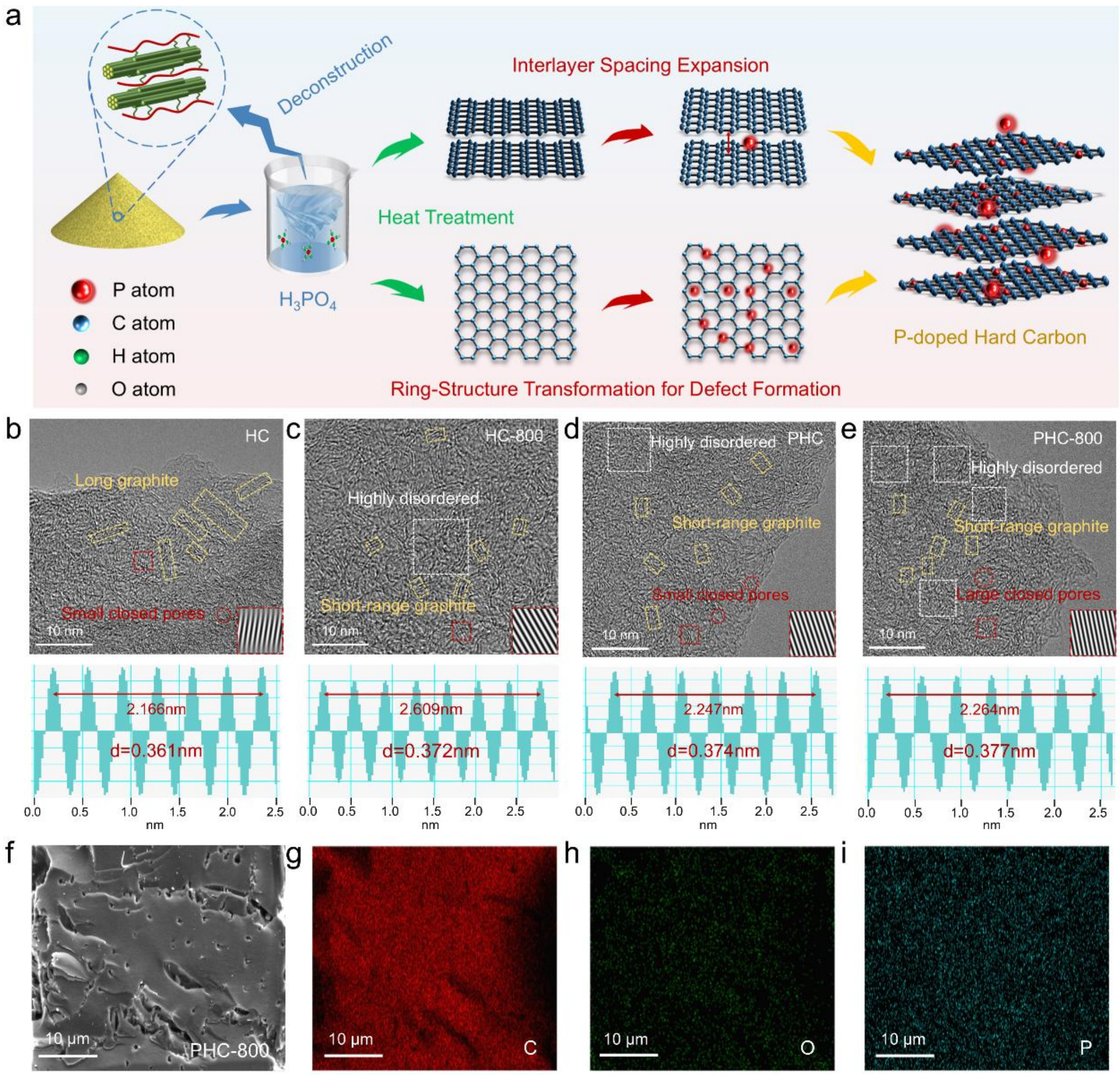

**Figure 1 Synthesis route of hard carbon and corresponding morphological characterization.** (a) Schematic illustration of the precursor structure. TEM images

and corresponding interlayer spacings of (b) HC, (c) HC-800, (d) PHC, and (e) PHC-800. (f) FIB-SEM image of PHC-800 and the corresponding elemental maps of (g) C, (h) O, and (i) P.

To investigate the structural evolution of hard carbon, X-ray diffraction (XRD) patterns of HC, HC-800, PHC, and PHC-800 were collected, as shown in Fig. 2a. Two broad diffraction peaks are observed at approximately 23° and 43°, corresponding to the (002) and (100) planes of amorphous carbon, respectively.[13, 17, 19] Notably, compared with HC, HC-800, and PHC, the (002) peak of PHC-800 shifts to lower angles. Based on Bragg's law, the average interlayer spacing of PHC-800 (0.376 nm) is larger than that of PHC (0.371 nm), HC-800 (0.370 nm), and HC (0.361 nm) (Table S2), indicating a significant interlayer expansion in PHC-800. These results are consistent with the TEM-measured lattice spacing (Fig. S2), confirming that the P doping and MT treatment increase interlayer spacing, thereby enhancing $Na^+$ intercalation kinetics. From the XRD data, the evolution of microcrystallite dimensions, $L_a$ (in-plane) and $L_c$ (stacking height), for HC, HC-800, PHC, and PHC-800 was calculated and is summarized in Fig. 2b. It is found that the $L_a$ decreases continuously from 4.622 nm for HC to 4.316 nm for PHC-800. Simultaneously, $L_c$ slightly recovers to 1.080 nm in PHC-800. This behavior can be attributed to the introduction of P, which induces local structural distortions that disrupt the in-plane periodicity of graphitic microcrystals, resulting in a continuous decrease in $L_a$. Increased defect concentration also hinders the ordered stacking of carbon layers, leading to an initial reduction in $L_c$.[20] The MT treatment provides limited atomic mobility, allowing local reorientation and rearrangement of carbon layers, converting portions of the disordered structure into short-range-ordered pseudo-graphitic domains and further compressing the in-plane crystalline size, thus driving the minimum $L_a$ value. Raman spectroscopy of HC, HC-800, PHC, and PHC-800 was also conducted (Fig. 2c). All hard carbons exhibit prominent D and G bands, with the intensity ratio ($I_D/I_G$) serving as a valuable parameter for evaluating defect concentration and degree of graphitization in carbon materials.[21] Deconvolution of the two main peaks reveals four sub-peaks of D (1345 $cm^{-1}$), D4 (1200 $cm^{-1}$), D3 (1500 $cm^{-1}$), and G (1590 $cm^{-1}$). Compared with other hard carbons, PHC-800 shows an increase in $I_D/I_G$ from 1.15 to 1.31, indicating an increased structural disorder within the carbon matrix (Fig. 2d and Fig. S3). These observations collectively demonstrate that P doping, combined with MT treatment, promotes the formation of a more disordered carbon structure at high temperatures.

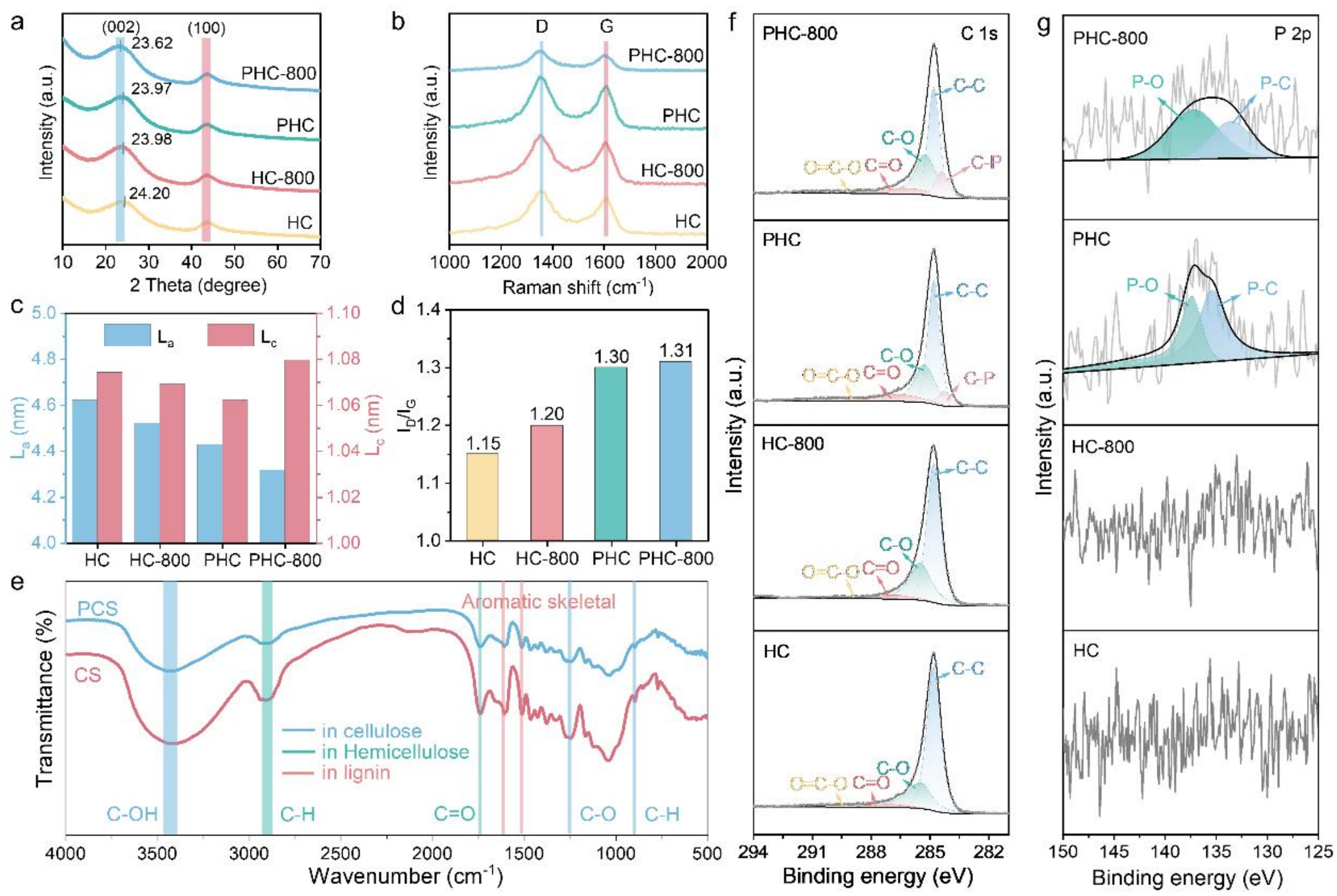

**Figure 2 Structural and interfacial characterization of hard carbon.** (a) XRD patterns and (b) corresponding microcrystallite parameters ($L_a$ and $L_c$) of HC, HC-800, PHC, and PHC-800. (c) Raman spectra and (d) corresponding $I_D/I_G$ ratios of HC, HC-800, PHC, and PHC-800. (e) FTIR spectra of CS and PCS. (f) C 1s and (g) P 2p XPS spectra of HC, HC-800, PHC, and PHC-800.

To elucidate the effect of phosphoric acid on the biomass precursor, Fourier-transform infrared (FTIR) spectroscopy was employed to monitor structural evolution. As shown in Fig. 2e, the absorption band at 3440 $cm^{-1}$ appears weaker in PCS compared to CS, likely due to the breakdown of crystalline microfibrils into glucose monomers and other small molecules during phosphoric acid pre-treatment.[22] Slight differences are also seen at 1730 $cm^{-1}$, associated with C=O stretching in hemicellulose, and at 900 $cm^{-1}$, related to C-H bending in cellulose, indicating minor structural variations between PCS and CS. The peaks of aromatic skeletal vibrations at 1602 and 1510 $cm^{-1}$ show decreased intensity in PCS in lignin, implying that phosphoric acid partially decomposes lignin and hemicellulose, which reduces the degree of crosslinking.[23] X-ray photoelectron spectroscopy (XPS) was further employed to investigate the influence of P doping on chemical bonding. The survey XPS spectra of PHC-800, PHC, HC-800, and HC are shown in Fig. S5. Notably, the P 2p signal is detected only in PHC-800 and PHC, confirming successful P incorporation. High-resolution C 1s, O 1s, and P 2p spectra were deconvoluted to further analyze bonding states (Fig. 2f, 2g, and Fig. S4). In the C 1s spectra (Fig. 2f), PHC-800 and PHC display a main peak at 284.8 eV corresponding to C-C bonds, indicating that the carbon framework remains predominantly composed of C-C interactions. Both PHC-800 and PHC exhibit a shoulder peak at 284.0 eV, attributed to C-P bonds, thereby further confirming P incorporation. In the O 1s spectra (Fig. S4), PHC-800 and PHC display characteristic O-P peaks at 531.4 eV, demonstrating that P introduction alters the surface chemical environment of the carbon materials. These findings are further corroborated by P 2p spectra (Fig. 2g), which reveal two components in the doped samples: P-O, indicative of P in oxidized states, and P-C, confirming direct chemical bonding between P and the carbon framework. In contrast, no discernible P 2p signals are observed for HC-800 and HC, indicating the absence of P incorporation in these samples.

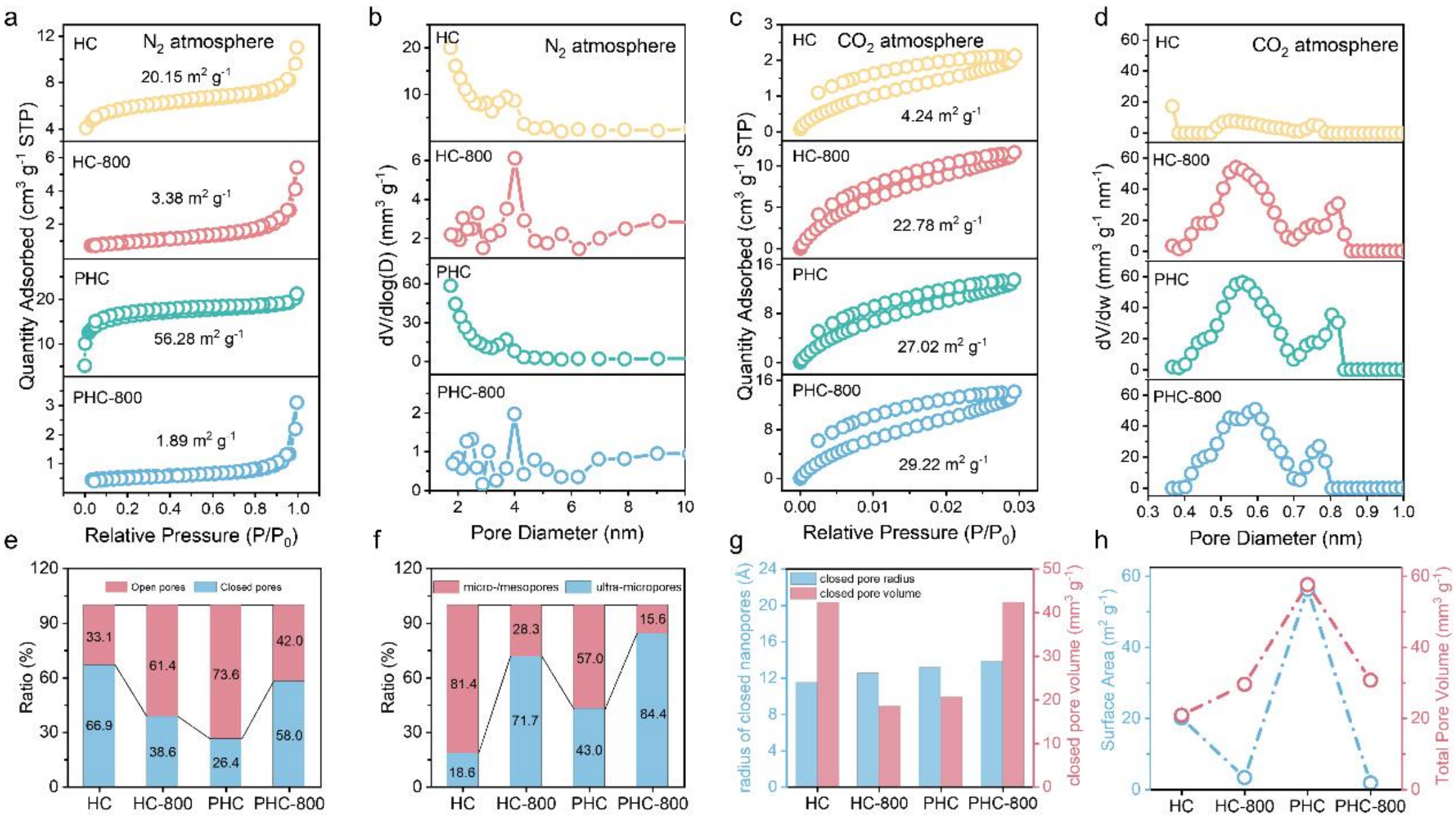

**Figure 3 Porous structure of hard carbon.** (a) $N_2$ adsorption-desorption isotherms and (b) corresponding pore size distributions of HC, HC-800, PHC, and PHC-800. (c) $CO_2$ adsorption-desorption isotherms and (d) corresponding pore size distributions derived of HC, HC-800, PHC, and PHC-800. (e) Volume fractions of closed and open pores, (f) volume fractions of ultramicropores and micro-/mesopores, (g) closed-pore radii and closed-pore volumes, and (h) changes in specific surface area and total pore volume of HC, HC-800, PHC, and PHC-800.

To further investigate the evolution of hard carbon porosity, $N_2$ adsorption-desorption isotherms were conducted on HC, HC-800, PHC, and PHC-800. The corresponding $N_2$ adsorption-desorption isotherms are shown in Fig. 3a. The BET specific surface area of PHC-800 is 1.89 $m^2 g^{-1}$, significantly lower than that of the other samples. Notably, the surface area of HC-800 (3.38 $m^2 g^{-1}$) is markedly reduced compared with HC (20.15 $m^2 g^{-1}$), indicating that MT treatment induces pre-reconstruction of the carbon framework. This facilitates local orientation and stacking of carbon layers, causing partial initial open pores to collapse or convert into closed pores during subsequent high-temperature processing, resulting in a substantial reduction in surface area. In contrast, PHC exhibits a significantly higher surface area (56.28 $m^2 g^{-1}$) than HC, suggesting that phosphoric acid acts as a chemical activating agent during thermal treatment. The extremely low surface area of PHC-800 arises from the synergistic effect of P doping and MT treatment: the carbon framework undergoes structural rearrangement supported by P atoms, converting many initial open pores into closed pores during high-temperature processing. While $N_2$ adsorption is primarily sensitive to pores >1 nm, $CO_2$ adsorption can probe ultramicropores <1 nm. $CO_2$ adsorption-desorption analysis (Fig. 3c) indicates that PHC-800 possesses a slightly higher surface area in the ultramicropore region. Pore size distribution derived by $CO_2$ adsorption-desorption (Fig. 3d) shows that all samples contain ultramicropores in the 0.4-0.8 nm range. Total open pore volumes for PHC-800, PHC, HC-800, and HC are 0.0307, 0.0576, 0.0296, and 0.0209 $cm^3 g^{-1}$, respectively (Table S4), whereas the open pore volumes for pores <1 nm are 0.0259, 0.0248, 0.0212, and 0.0039 $cm^3 g^{-1}$, respectively. Notably, ultramicropores (<1 nm) account for 84.4% of the open pore volume in PHC-800, significantly higher than HC (18.6%), HC-800 (71.7%), and PHC (43.0%). These results indicate that MT treatment significantly promotes ultramicropore formation. Under the synergistic effect of P doping and MT treatment, the carbon framework preferentially forms pores <1 nm, thereby enhancing $Na^+$-carbon wall interactions and improving adsorption stability.[24-25]

To evaluate closed-pore characteristics, small-angle X-ray scattering (SAXS) was employed. As shown in Fig. S6, within the $q$ range of 0.01-0.5 $Å^{-1}$, PHC-800 exhibits higher scattering intensity. SAXS fitting indicates that the closed-pore radii of PHC-800 (1.38 nm) are larger than those of PHC (1.32 nm), HC-800 (1.25 nm), and HC (1.15 nm), consistent with TEM observations. Helium pycnometry was further used to determine the true density and evaluate closed-pore volume, calculated as[26]

$$V_{closed} = \frac{1}{\rho_{true}} - \frac{1}{2.26}$$

Based on measured true densities (Table S3), the closed-pore volumes of PHC-800, PHC, HC-800, and HC are 0.0424, 0.0207, 0.0186, and 0.0423 $cm^3\ g^{-1}$, respectively. Interestingly, HC-800, despite MT treatment, exhibits a reduced closed-pore volume, suggesting that MT treatment primarily promotes ultramicropore formation rather than efficient open-to-closed pore conversion. PHC, although P-doped, retains a relatively low closed-pore volume, indicating that chemical activation mainly generates open pores. Only the combination of P doping and MT treatment achieves simultaneous maximization of both closed-pore radius and volume, in which P initially establishes stable phosphate-ester bridging structures that are subsequently converted into closed pores at high temperature. The synergistic MT effect further facilitates ultramicropore formation, yielding PHC-800 with high concentrations of both closed and ultramicropores, enhancing Na-ion storage. A schematic illustrating how these features contribute to performance improvement is presented in Fig. S7.

Electrochemical characterization was performed to correlate carbon structure with performance. Fig. 4a presents the initial galvanostatic charge-discharge (GCD) profiles at 20 $mA\ g^{-1}$ and corresponding ICEs for PHC-800, PHC, HC-800, and HC electrodes. It is found that the PHC-800 exhibits the highest ICE (90.4%), exceeding PHC (89.6%), HC-800 (87.6%), and HC (83.1%). A clear negative correlation between ICE and surface area is observed. Lower surface area reduces electrolyte exposure of open pores, minimizing irreversible capacity loss due to SEI formation. PHC-800, with the lowest surface area (1.89 $m^2$ $g^{-1}$), shows the highest ICE, whereas PHC (56.28 $m^2$ $g^{-1}$) shows a slightly lower ICE. A similar trend is observed for HC-800 versus HC (3.38 $m^2$ $g^{-1}$ vs. 20.15 $m^2$ $g^{-1}$; 87.6% vs. 83.1%). Rate capability measurements (Fig. 4b) reveal that PHC-800 delivers the highest reversible capacity of 342.3 mAh $g^{-1}$, whereas PHC, HC-800, and HC provide 326.0, 306.1, and 277.4 mAh $g^{-1}$, respectively. PHC-800 also demonstrates excellent reversibility, retaining 338.8 mAh $g^{-1}$ when the current returns to 20 mA $g^{-1}$. All hard carbons display typical electrochemical behavior, with sloping regions above 0.1 V and plateau regions below 0.1 V.[27] Fig. 4c shows the capacity contribution from sloping and plateau regions during the second cycle. HC, HC-800, and PHC exhibit limited plateau capacities of 199.2, 225.8, and 248.7 mAh $g^{-1}$, respectively, whereas PHC-800 achieves the highest plateau capacity of 262.3 mAh $g^{-1}$, corresponding to 76.6% of its reversible capacity. The expansion of the plateau region is attributed to the abundant ultramicropores, which act as additional $Na^+$ storage sites and facilitate ion transport, thereby enhancing plateau capacity and improving ICE.[28] Cycling performance at 20 mA $g^{-1}$ over 100 cycles (Fig. 4d) demonstrates near 100% capacity retention for all the hard carbons. High-current cycling of PHC-800 at 300 mA $g^{-1}$ for 1000 cycles shows 84.5% capacity retention (Fig. 4e), highlighting excellent stability. The superior electrochemical performance of PHC-800 is attributed to the coupled effects of carbon layer reorientation and closed-pore construction.

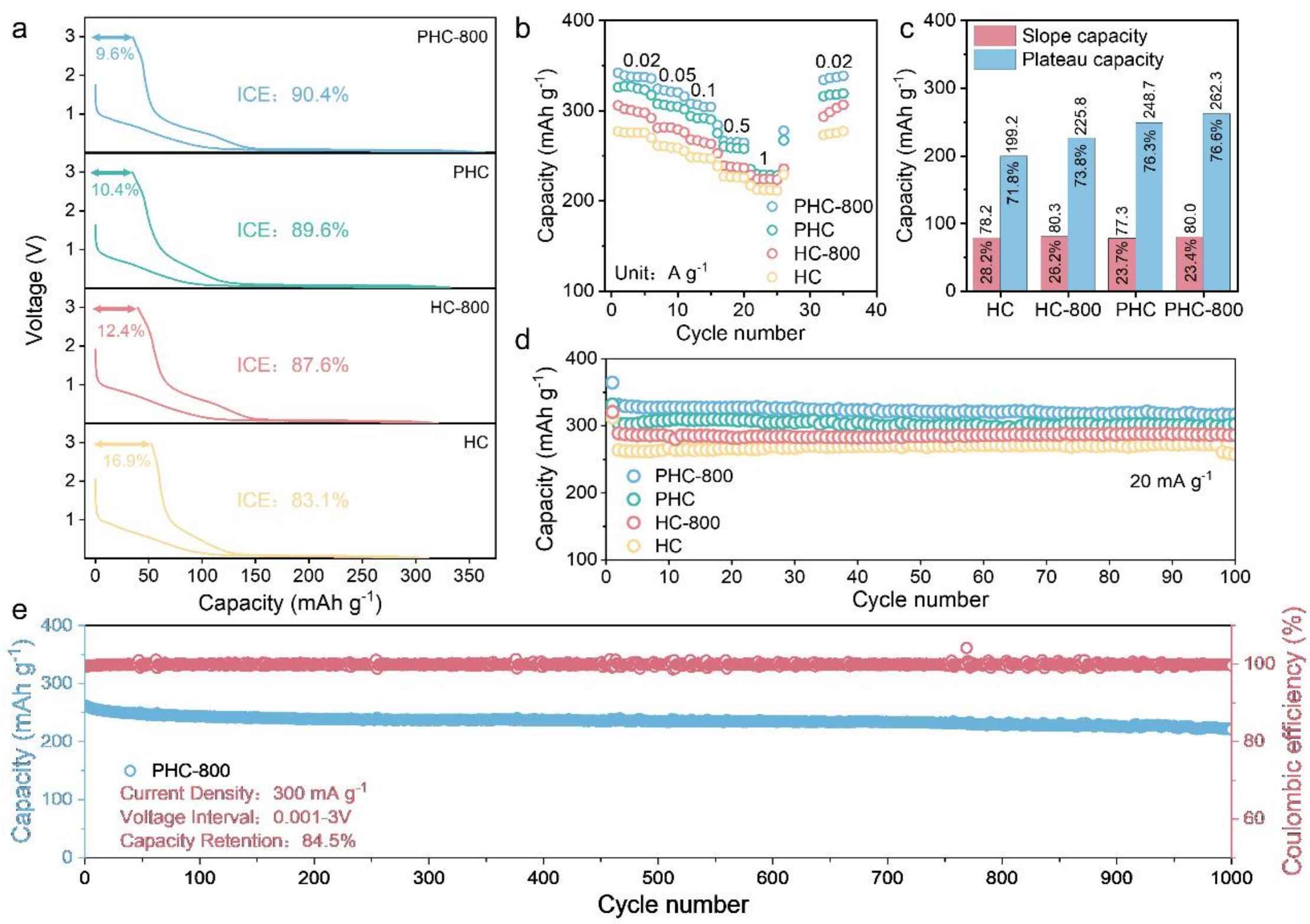

**Figure 4 Na-ion storage performance of hard carbon.** (a) Initial GCD profiles, (b) rate capability, and (c) sloping- and plateau-region capacities and (d) cycling performance for HC, HC-800, PHC, and PHC-800. (e) Long-term cycling performance of PHC-800 over 1000 cycles at 300 mA $g^{-1}$.

Cyclic voltammetry (CV) at various scan rates was performed to investigate the charge storage mechanisms of the four electrodes (Fig. S9). Well-defined redox peaks are observed even at high scan rates, indicating low polarization and excellent stability of the constructed electrodes. The *b*-values, derived from the power-law relationship (Fig. S10):

$$i = av^b$$

Here, *i* represents the peak current and *v* denotes the scan rate. Based on the calculation, PHC-800 exhibits a relatively high b-value of 0.87., suggesting a dominant surface-controlled behavior.[29] Fig. S12 shows the CV curves of PHC-800, PHC, HC-800, and HC at a scan rate of 0.1 mV $s^{-1}$. In the initial cycle, a broad irreversible redox peak at ~0.8 V is observed, attributed to electrolyte decomposition and the formation of the SEI on the electrode surface. Samples with larger open pore surface areas, such as PHC and HC, display more pronounced irreversible peaks due to excessive SEI formation and irreversible $Na^+$ adsorption at defect sites. In contrast, PHC-800 exhibits a significantly smaller irreversible peak area, indicating that the coupled effect of P doping and MT treatment effectively reduces irreversible $Na^+$ consumption. Subsequent cycles show highly overlapping CV curves for all electrodes, demonstrating good $Na^+$ reversibility. As shown in Figs. S13-S16, the capacitive and diffusion-controlled contributions were quantified by separating current responses.[30] According to the calculation, the capacitive contribution of PHC-800 reaches 64.7% at 0.1 mV $s^{-1}$, compared with 19.4%, 38.2%, and 69.0% for PHC, HC-800, and HC, respectively. Increasing the scan rate increases the capacitive contributions for all electrodes (Fig. S11). Electrochemical impedance spectroscopy (EIS) coupled with distribution of relaxation times (DRT) analysis under different potentials (OCP-0.01 V) was conducted to probe the potential-dependent interfacial and charge-transfer kinetics during discharge (Figs. 5e-h).[31] Notably, during discharge from open-circuit potential to 0.8 V, the SEI impedance increases for all four electrodes, primarily due to the initial reduction of the electrolyte at the electrode surface. The nascent SEI layer, dominated by organic components, is relatively loose with low ionic conductivity, thereby hindering $Na^+$ transport at the interface.[32-33] As the potential further decreases to 0.01 V, the SEI impedance diminishes gradually. This trend results from the structural reorganization of the SEI, where conductive inorganic components such as NaF and $Na_3PO_4$ progressively replace the original organic layer, thereby improving interfacial ion transport. Additionally, sealed pores not directly exposed to the electrolyte do not contribute to SEI formation. Pore filling also alleviates internal electrode stress, which helps stabilize the SEI.[34] Furthermore, $Na^+$ intercalation into both the carbon interlayers and closed pores at low potentials induces electrochemical activation, further enhancing ion transport pathways.

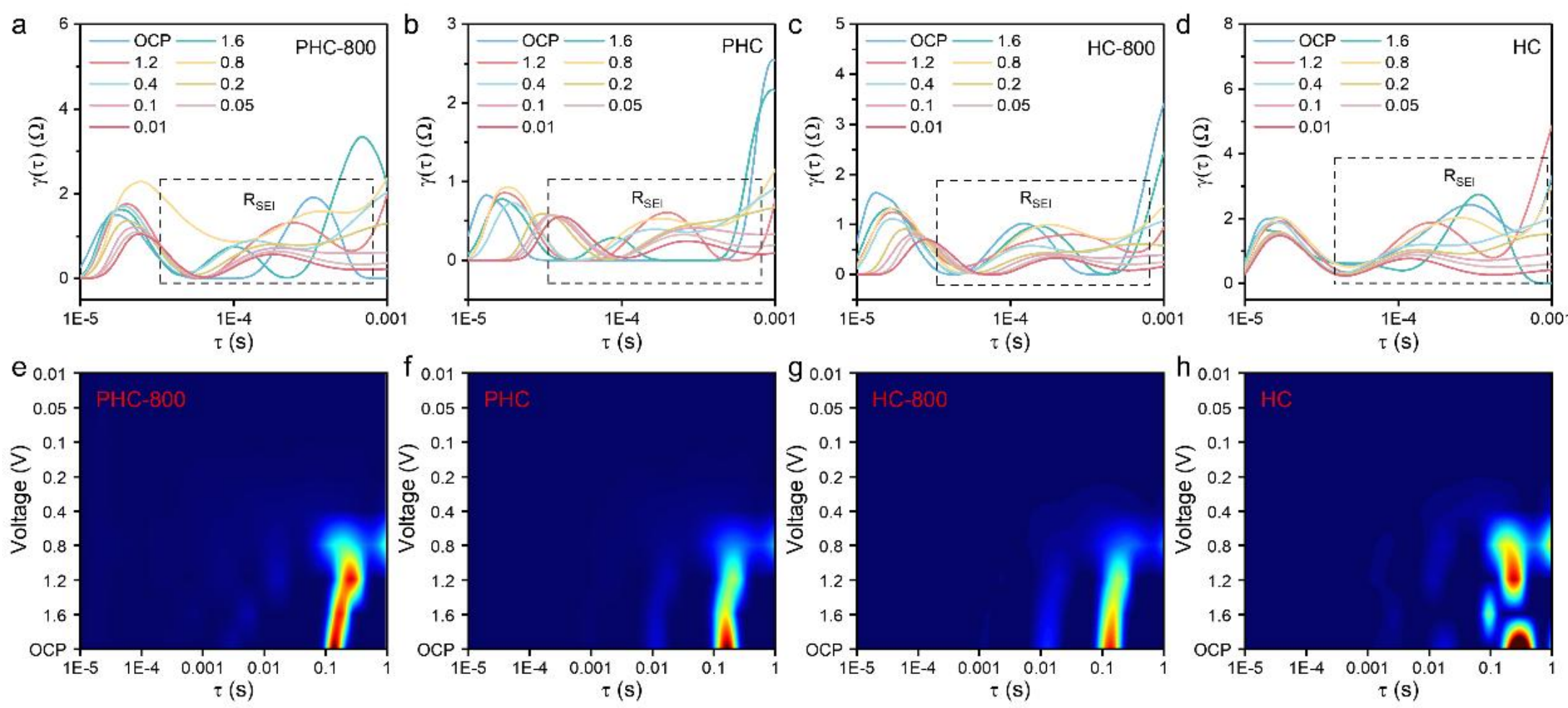

**Figure 5 DRT characteristics.** (a-d) DRT spectra of (a) PHC-800, (b) PHC, (c) HC-800 and (d) HC in the high-frequency region. Contour plots of DRT spectra of (e) PHC-800, (f) PHC, (g) HC-800 and (h) HC.

To further elucidate the Na-ion storage mechanism in hard carbon electrodes, the galvanostatic intermittent titration technique (GITT) was employed to study the dynamic evolution of $Na^+$ diffusion coefficients during charge-discharge (Fig. S18 and Fig. 6a). The $Na^+$ diffusion coefficients for all four electrodes exhibit highly consistent trends, indicating similar Na-ion storage kinetics. Distinct variations in diffusion coefficients are observed as the storage mechanism transitions from surface adsorption to interlayer insertion.[35] From open-circuit potential to 0.78 V, the diffusion coefficient increases, attributed to initial electrolyte reduction at the electrode surface, forming a loose organic SEI layer with low ionic conductivity.[36] The diffusion coefficient subsequently decreases, corresponding to SEI formation at defect sites via $Na^+$ adsorption. When the voltage decreases further to 0.1 V, $Na^+$ diffusion coefficients increase again. The dynamic migration of $Na^+$ within hard carbon in the plateau region indicates an early-stage migration from interlayer/adsorption sites to quasi-metallic Na clusters in closed pores, followed by a synergistic adsorption/interlayer occupation process. Notably, adsorbed $Na^+$ exhibits higher mobility than interlayer-inserted $Na^+$, accompanied by a temporary decrease in diffusion coefficient.[37] Within the voltage range of 0.1-0.04 V, diffusion coefficients decrease sharply, reflecting the transition from surface adsorption to interlayer insertion and closed-pore filling, where $Na^+$ accumulates at transition sites (graphitic microcrystal edges, <1 nm), increasing electrostatic repulsion and hindering diffusion.[38] As the voltage further decreases from 0.04 V to 0 V, diffusion coefficients rise rapidly, consistent with pore-filling processes. Combined with BET and pore size analyses, these results suggest that PHC-800 possesses a pore size distribution optimally suited for $Na^+$ accommodation, highlighting the advantage of the coupled mechanism of carbon layer reorientation and closed-pore construction.

To correlate structural evolution with Na-ion storage, in situ Raman spectroscopy was performed to monitor changes during charge-discharge (Fig. 6b-d). In the high-voltage region, the D and G bands of PHC-800 show negligible shifts, indicating that $Na^+$ adsorption at defect sites minimally perturbs the graphitic lattice. During low-voltage discharge, the D-band position remains unchanged, while its intensity decreases significantly due to suppressed breathing vibrations of the graphitic lattice at pore edges.[39] The G-band exhibits a pronounced intensity reduction and redshift, associated with electron-phonon coupling induced by $Na^+$-carbon interactions.[40] The G-band position shift reflects progressive $Na^+$ insertion and uniform diffusion within each carbon layer, resulting in electron doping.[41-42] Additionally, G-band broadening in the low-voltage region indicates that $Na^+$ diffusion within pore structures weakens C-C bond strength, further corroborating the intrinsic correlation between material structure and Na-ion storage performance. Based on these observations, the Na-ion storage mechanism of PHC-800 can be described by an adsorption-intercalation-pore filling model (Fig. 6e). To further probe Na-ion storage in the plateau region, PHC-800 electrodes were discharged to different voltages, removed, and immersed in ethanol containing phenolphthalein. As shown in Fig. S19, the solution exhibits varying degrees of magenta coloration at voltages below 0.1 V, indicating the reaction of Na clusters stored in closed pores with ethanol to generate $H_2$ and $CH_3CH_2ONa$. The resulting alkalinity induces phenolphthalein color changes. As the voltage decreases to 0.001 V, the solution color deepens, indicating increased Na cluster storage in closed pores. Conversely, upon increasing the voltage, the color lightens, confirming the reversibility of Na cluster formation and dissolution.[43] The phenolphthalein experiment thus provides further evidence for the adsorption-intercalation-pore filling mechanism in PHC-800, offering critical insight into hard carbon Na-ion storage and highlighting the importance of appropriately sized closed and ultramicropores for enhancing plateau-region capacity.

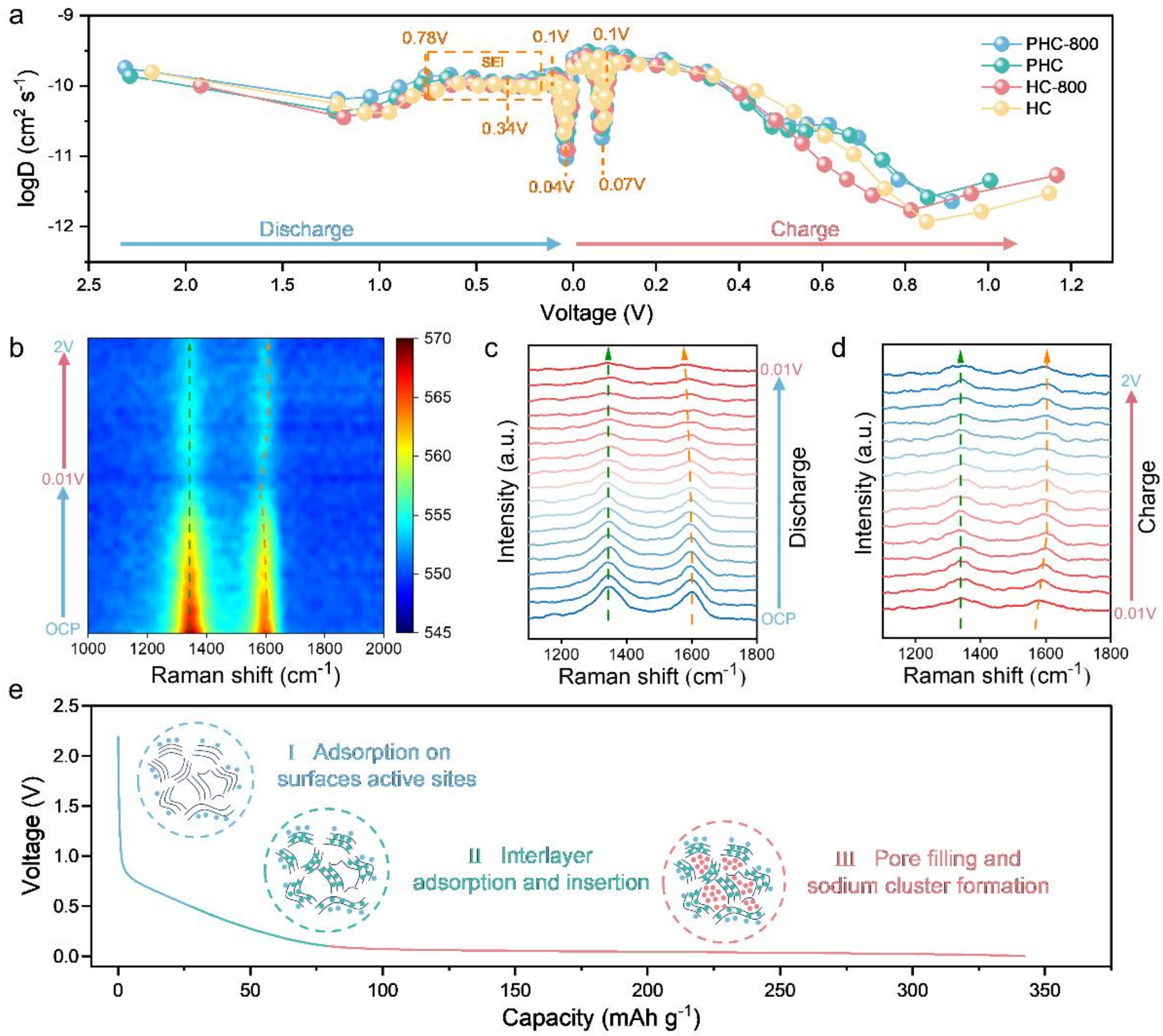

**Figure 6 Na-ion diffusion and in situ Raman analysis**. (a) Calculated $Na^+$ diffusion coefficients of PHC-800, PHC, HC-800 and HC during sodiation-desodiation process. (b) in situ 2D Raman spectra of PHC-800 during charge-discharge process. In situ Raman spectra during (c) discharge and (d) charge process. (e) Schematic illustration of the Na-ion storage stages in hard carbon.

To evaluate practical performance, a full coin cell was assembled using PHC-800 as the anode and $Na_3V_2(PO_4)_3$ (NVP) as the cathode (Fig. 7a). It is well established that an appropriate N/P ratio is critical for both battery safety and performance, where N/P is defined as the ratio of the total reversible capacity of the anode to that of the cathode.[44] To prevent Na dendrite formation and excessive consumption of metallic Na-ion during charging, the full cell was designed with an N/P ratio >1. As shown in Fig. 7b, the full cell was tested within a voltage window of 2.2-3.8 V. After 70 cycles at 0.1 A $g^{-1}$, a capacity retention of 98.1% was achieved, demonstrating excellent stability. Rate performance tests indicate reversible capacities of 103.4, 100.7, 98.0, 93.9, and 82.0 mAh $g^{-1}$ at current densities of 0.05, 0.1, 0.2, 0.5, and 1.0 A $g^{-1}$, respectively. Notably, when the current density was returned to 0.05 A $g^{-1}$, the cell recovered a fully reversible capacity of 97.39 mAh $g^{-1}$, highlighting outstanding electrochemical reversibility. These results, together with the high-rate performance shown in Fig. 7c-d, confirm the practical feasibility of PHC-800 hard carbon as an anode material for sodium-ion batteries.

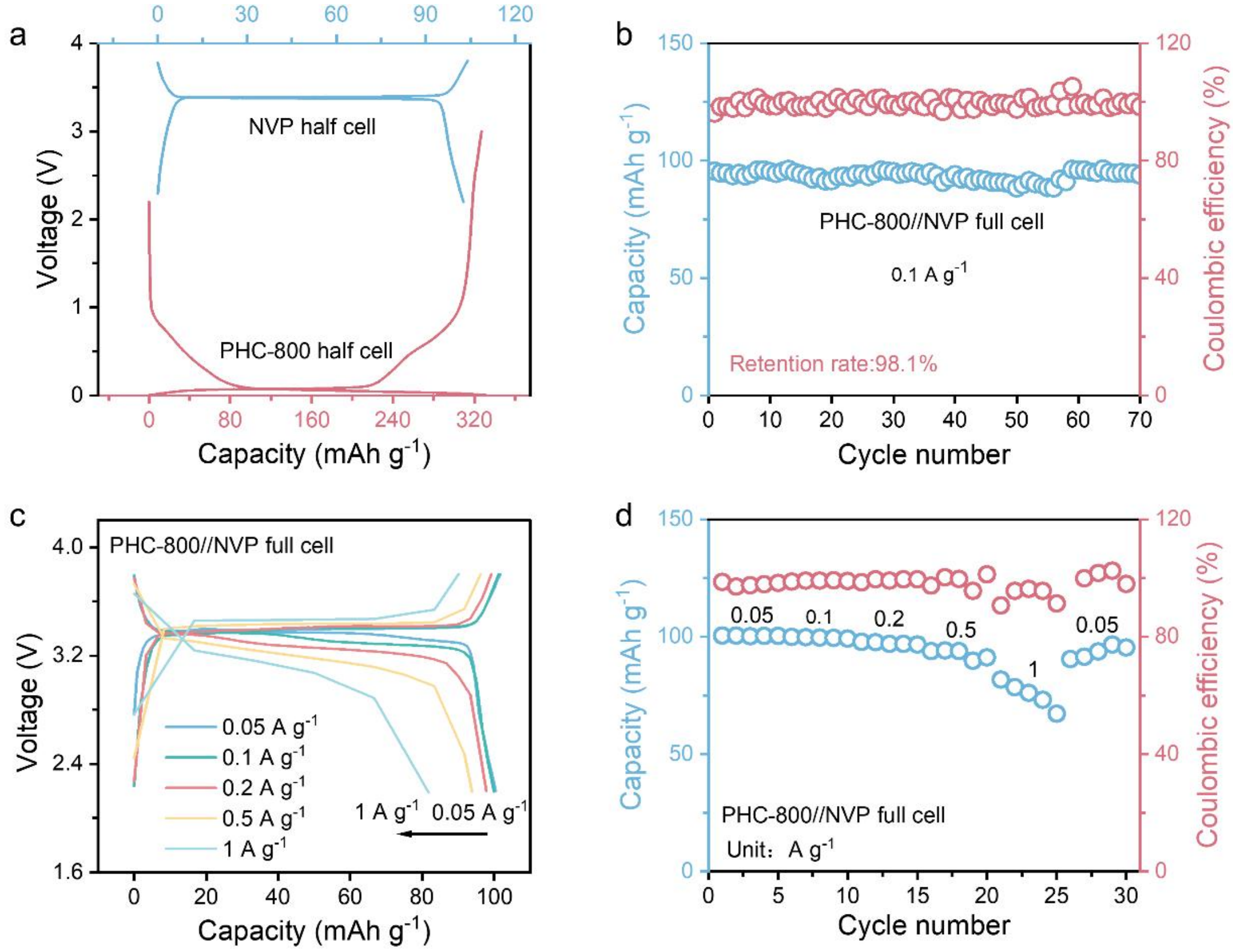

**Figure 7 Full-cell performance.** (a) Capacity matching between PHC-800 anode and NVP cathode. (b) Cycling performance of the PHC-800//NVP full cell at 0.1 A $g^{-1}$. (c) Cycling performance and (d) rate capability of the PHC-800//NVP full cell.

## 3. Conclusion

In summary, we propose a carbon layer reorientation coupled closed-pore construction to guide the structural design of high-performance hard carbon anodes for NIBs. Leveraging P doping and MT thermal treatment, we found that our hard carbon exhibits enhanced structural disorder and directional formation of closed/ultramicropore networks. The resulting PHC-800 hard carbon exhibits a low specific surface area (1.89 $m^2 g^{-1}$), high closed-pore volume (0.0424 $cm^3 g^{-1}$) with abundant ultramicropores, which leads to a high ICE (90.4%) and a plateau-region capacity of 262.3 mAh $g^{-1}$ (76.6% of the reversible capacity) for Na-ion storage. We also reveal that the Na-ion storage mechanism follows an adsorption-intercalation-pore-filling model, as confirmed by electrochemical and structural analyses, thereby confirming the effectiveness of the coupled strategy in simultaneously enhancing capacity and first-cycle efficiency. We believe that our work provides a viable pathway for designing low-surface-area hard carbon anodes with high ICE for advanced Na-ion storage

**Acknowledgments**

We thank the financial support from the Guangxi Science and Technology Program (2024AB08156), the Natural Science Foundation of Guangxi Province (2024GXNSFBA010183), the Program for Bagui Scholars of Guangxi, China, and the Innovation Project of GUET Graduate Education (2026YCXS180).

**Conflict of Interest**

The authors declare no conflict of interest.